\documentclass[aps,pra,twocolumn,superscriptaddress, 10pt,longbibliography]{revtex4-1}
\usepackage{mathtools}
\usepackage{amsmath,amscd,amssymb,amsthm}
\usepackage{xcolor}
\begin{document}

\title{Single-photon description of the lossless optical Y-coupler}
\author{Christopher R. Schwarze}
\email[e-mail: ]{{\tt crs2@bu.edu}}
\affiliation{Department of Electrical and Computer Engineering \& Photonics Center, Boston University, 8 Saint Mary’s St., Boston, Massachusetts 02215, USA}
\author{Anthony D. Manni}
\email[e-mail: ]{{\tt admanni@bu.edu}}
\affiliation{Department of Electrical and Computer Engineering \& Photonics Center, Boston University, 8 Saint Mary’s St., Boston, Massachusetts 02215, USA}
\author{David S. Simon}
\email[e-mail: ]{{\tt simond@bu.edu}}
\affiliation{Department of Electrical and Computer Engineering \& Photonics Center, Boston University, 8 Saint Mary’s St., Boston, Massachusetts 02215, USA}
\affiliation{Department of Physics and Astronomy, Stonehill College, 320 Washington Street, Easton, Massachusetts 02357, USA}
\author{Alexander V. Sergienko}
\email[e-mail: ]{{\tt alexserg@bu.edu}}
\affiliation{Department of Electrical and Computer Engineering \& Photonics Center, Boston University, 8 Saint Mary’s St., Boston, Massachusetts 02215, USA}
\affiliation{Department of Physics, Boston University, 590 Commonwealth Avenue, Boston, Massachusetts 02215, USA}

\date{August 27, 2024}
\begin{abstract}
Using symmetry considerations, we derive a unitary scattering matrix for a three-port optical Y-coupler or Y-branch. The result is shown to be unique up to external phase shifts. Unlike traditional passive linear-optical one-way splitters, coupling light into the conventional output ports of the Y-coupler results in strong coherent back-reflections, making the device a hybrid between feed-forward devices like the beam-splitter, which do not reverse the direction of light, and a recently considered class of directionally-unbiased multiport scatterers (with dimension greater than two) which do. While the device could immediately find use as a novel scattering vertex for the implementation of quantum walks, we also design a few simple but nonetheless useful optical systems that can be constructed by taking advantage of the symmetry of the scattering process. This includes an interference-free, resource-efficient implementation of the Grover four-port and a higher-dimensional Fabry-Perot interferometer with tunable finesse. Symmetry-breaking generalizations are also considered.
\end{abstract}

\maketitle
\section{Introduction}
Symmetry has long been a critical tool for defining physical theories. With the constraints that symmetries carry, dimensionality and degrees of freedom can be reduced, simplifying the study of many systems which otherwise cannot be understood analytically. One instance where this tool is useful is in classifying linear-optical scattering transformations. As a simple example, quantum-mechanical $U(1)$ gauge symmetry allows any scattering matrices related by a global phase factor $e^{i\theta}$ to be considered equivalent because such a global phase shift is not measurable. Other physical symmetries in a single-photon scattering process translate to mathematical constraints on the corresponding scattering matrix, allowing scatterers to be designed within the subspace defined by these constraints. The constraints carried by symmetric scatterers are also often useful for simplifying the design and analysis of larger systems formed from many symmetric scatterers. 

The traditional view of feed-forward optical scattering treats the beam-splitter as a device with two input and two output ports. However, there is no physical distinction between a port treated as input or output; all ports are tied to a pair of counter-propagating field modes. Dropping the distinction between input and output allows a more general, directionally-unbiased scattering framework to be used, which we employ here. In this framework, a four-port beam-splitter would be given a $4\times 4$ scattering matrix. 

A Y-coupler or Y-branch is traditionally used to split one beam of light into two. It is distinct from a beam-splitter since it has three ports instead of four. There is only one traditional input port, so it cannot overlap two beams, but in situations where this is unnecessary, a passive Y-branch can be preferable over a beam-splitter for reasons such as compactness. Y-coupler designs are ubiquitous, as are their experimental demonstrations in waveguide systems \cite{Lin:19, Han:18, Zhang:13, 19110, 4840618, 948285, Wang:02,10.1063/5.0011452, Opielka1979, Bogert1989, 350602, Cooney:16}. The device has also been realized in photonic-crystals \cite{Chantakit2014}, and can be readily constructed using bulk materials, which we will describe in further detail later. There have also been some proposals and demonstrations to employ one or more low-loss Y-couplers in larger systems \cite{Liu2015, PhysRevLett.113.103601, 531823, Miller:20} such as for coherent applications like interferometry \cite{Yadav:22, app13063829, holmenthesis}, heterodyning \cite{4305230}, and amplification \cite{Uemukai2012IntegratedTD, Sanford:91}.

Many Y-couplers have been designed and realized for use as a compact beam-divider. When the Y-coupler input is confined to only the traditional input port, it's description is incomplete in our general, directionally-unbiased scattering formalism. A complete description of the device requires understanding its response to light being input separately to each of its traditional output ports.

In this work, we use an assumed set of symmetries to derive the full unitary scattering matrix for the standard optical Y-coupler. We later show this leads to major differences in the behavior of this device in comparison to a traditional beam-splitter. For instance, light entering a beam-splitter cannot reverse direction, but this is not the case for the Y-coupler. The main result is that this well-known Y-coupler is truly a hybrid scatterer which is directionally-unbiased or feed-forward depending on the input port. The unitary scattering matrix derived here provides this complete description of the device and consequently suggests a number of interesting systems can now be formed using the directionally-biased ports in conjunction with the traditional input port. A few examples are provided in Section IV. Among them is a realization of the four-port Grover coin which significantly improves the only other known optical implementation. This scatterer plays a key role in unrealized theoretical proposals for enhanced quantum optical interferometry and sensing \cite{PhysRevA.106.033706, PhysRevA.107.052615}, quantum information processing \cite{PhysRevA.104.012617, PhysRevA.104.012617}, quantum walks and quantum search \cite{PhysRevA.101.032118, PhysRevA.95.042109, PhysRevA.96.013858, kubota2022perfect, giri2017review}. Employing the directionally-unbiased ports of the Y-coupler allows construction of novel multiport interferometers with useful properties, such as the generalized Fabry-Perot resonator that will be discussed in detail in Section IV.

In the next section, we illustrate some important optical scattering process symmetries. After this, in Section III, we use a specific set of such symmetries to uniquely derive the field transformation for the Y-coupler. In Section V we discuss some potential applications of these devices and draw final conclusions. Generalizations that occur when symmetries are broken are analyzed in the appendix.

\section{Optical scattering symmetries}

The ideal model of many quantum electromagnetic scattering processes assumes time-reversal symmetry,  originating from Hermiticity of the underlying Hamiltonian $H$, and reciprocity, guaranteed whenever there exists a certain anti-unitary operator $K$ satisfying $KHK^{-1} = H^\dagger$ \cite{DEAK20121050}. These symmetries respectively imply unitarity $U = U^\dagger$ and diagonal symmetry $U = U^T$ of the scattering matrix.

Some scatterers also exhibit forms of geometric symmetry. For example, scattering structures which are invariant under point group rotations behave identically no matter which port is used for input. This implies the numerical labels assigned to each port can be cyclically permuted without affecting the form of the scattering matrix. The same permutation-symmetric structure is accordingly imposed on the scattering matrix: neighboring columns are neighboring cyclic permutations of each other. Such matrices are known as circulant, and have recently been studied in physical contexts, such as quantum walks \cite{PhysRevA.106.022402}. These matrices are well-known in numerical analysis due to the property that they are diagonalized by a discrete Fourier transform \cite{davis1979circulant}.

A prevalent circulant optical device is the circulator. Devices in this class are defined by the characteristic that light entering port $m$ always exits at port $(m + j) \mod N$ for some $m, j \in \{1, \dots, N\}$ such that $(m + j) \mod N \neq N/2$ when $N$ is even. This additional constraint is to ensure the devices in this class lack reciprocal symmetry. For example, if $N = 4$ and $j = 1$, the scattering matrix is given by
\begin{equation}
C = 
  \begin{pmatrix}
    0 & 0 & 0 & 1\\
    1 & 0 & 0 & 0\\
    0 & 1 & 0 & 0\\
    0 & 0 & 1 & 0\\    
  \end{pmatrix},
\end{equation}
but if $N = 4$ and $j = 2$, then the matrix would be symmetric. 

Scattering transformations can also possess reflection symmetry about some line, so that pairs of output modes on either side of this line have equal amplitudes; each pair will be comprised of the outgoing modes with equal and opposite $\mathbf{k}$-vector components normal to the reflection line. A common reflection invariance is with respect to the line defined by the the input mode propagation vector $\mathbf{k}_{\text{in}}$, as shown in Fig. \ref{fig:yc}. The presence of a reflection symmetry allows the port labels of any pair of mirrored modes to be swapped without changing the scattering matrix. 

The $d$-port Grover coin (where $d \geq 3$) is highly symmetric \cite{kempe}. It is unitary, reciprocal, circulant, and possesses reflection symmetry. Because the device is circulant, the reflection symmetry automatically applies to all input ports, and thus all transmitting amplitudes are equal. This structure allows the port labels to be assigned arbitrarily, and any permutation made to these labels will not affect the scattering matrix. These assumptions impose the following form on a $d$-port unitary:
\begin{equation}
U = 
\begin{pmatrix}
  r & t & t & t & \dots & t\\
  t & r & t & t & \dots & t\\
  t & t & r & t & \dots & t\\
  t & t & t & r & & \vdots \\
  \vdots & \vdots & \vdots& &\ddots & t\\
  t & t & t & \dots & t & r
\end{pmatrix}.
\end{equation}
Orthogonality implies
\begin{equation}\label{eq:gorth}
  r^*t + t^* r + (d-2)|t|^2 = 0, 
\end{equation}
while normalization requires
\begin{equation}\label{eq:gnorm}
  |r|^2 + (d-1)|t|^2 = 1. 
\end{equation}
We assume $|t|$ is nonzero otherwise the device is acting as a mirror. Then, by expressing $r = |r|e^{i\phi_r}$ and $t = |t|e^{i\phi_t}$ in polar coordinates, we may rewrite the first equation as $2|r|\cos(\phi_r - \phi_t)+ (d-2)|t| = 0$. If we select $\phi_r - \phi_t = \pi$, we arrive at the Grover coin, as this gives $|r| = (d/2-1)|t|$, which when placed in the normalization condition, yields $t = 2/d$ and $r = 2/d - 1$. An additional symmetry emerges when $d = 4$, causing the reflected and transmitted probabilities to coalesce at the value 1/4. This can be tied to a reflection symmetry with respect to the forward transmitted and back-reflected output modes.

In the analysis above, different values of $\phi_r - \phi_t$ generate valid scatterers whenever $\cos(\phi_r - \phi_t) \leq 0$. The Grover coin can be identified as the value admitting minimal reflectivity within this reciprocal, unitary and circulant space. To see this, we let $x = \cos(\phi_r - \phi_t)$ and combine Eq. (\ref{eq:gorth}) and (\ref{eq:gnorm}) to obtain the constraint between $x$ and $|r|^2$,
\begin{equation}
    |r|^2 x^2 = c_0^2 (1-|r|^2),
\end{equation}
where $c_0 =  (2-d)/(2\sqrt{d-1})$. Solving for $|r|^2$, we have 
\begin{equation}
    |r|^2 = \frac{c_0^2}{c_0^2 + x^2}.
\end{equation}
Now, $x^2 = \cos^2(\phi_r - \phi_t)$ lies in the set $[0, 1]$; for $x = 0$, $|r|^2 = 1$. Increasing $x^2$ by varying $x$ from $0$ to $-1$ decreases $|r|^2$ until it reaches its minimum value, 
\begin{equation}
    |r_{\text{min}}|^2 = \frac{c_0^2}{c_0^2 + 1} = \frac{(d-2)^2}{(d-2)^2 + 4(d-1)} = \frac{(d-2)^2}{d^2}, 
\end{equation}
corresponding to the $d$-port Grover coin. In general, the other values of $|r|^2$ in this set can be spanned by placing a mirror on any port of a $(d+1)$-port Grover coin and varying the phase acquired between the coin and the mirror \cite{PhysRevA.107.052615}.

A traditional lossless, reciprocal beam-splitter assumes a feed-forward symmetry. This is physically encapsulated by the fact that light entering any one of the four input ports only can emerge from two of the ports on the other side; it cannot reverse direction. Mathematically this results in a $U(2)$ device being embedded in a four-dimensional space, 
\begin{equation}
  B =
\begin{pmatrix}
  0 & 0 & r_1 & t_2\\
  0 & 0 & t_1 & r_2\\
  r_1 & t_2 & 0 & 0\\
  t_1 & r_2 & 0 & 0
\end{pmatrix}.
\end{equation}
The symmetry can be expressed as an invariance of this matrix under the port label mapping $(1, 2) \longleftrightarrow (3, 4)$.

Because the device operates in a two-dimensional unitary subspace, there is little lost by assuming reciprocity within this subspace as well. This is because for general, non-reciprocal $U(2)$ devices, the unitarity constraint carries two implications. First, it requires $|r_1| = |r_2| = \sqrt{1 - |t_1|} = \sqrt{1 - |t_2|}$ so that non-reciprocity only can appear in the acquired phases. The second implication is that the phases must satisfy the constraint $\text{arg } r_1 + \text{arg } r_2 = \text{arg } t_1 + \text{arg } t_2 + \pi$.

In fact, the Hong-Ou-Mandel effect originates directly as a result of these phase and reciprocal probability constraints when the auxiliary assumption $|r_1| = 1/\sqrt{2}$ is made. For, the initial state of two indistinguishable monochromatic photons impinging adjacent ports of a beam-splitter will transform to a state
\begin{align}
   |\psi\rangle = (|r_1| e^{i\text{arg } r_1} a^\dagger + |t_1| e^{i\text{arg } t_1} b^\dagger) \\ \nonumber (|r_2| e^{i\text{arg } r_2} b^\dagger + |t_2| e^{i\text{arg }t_2} a^\dagger)|0\rangle
\end{align}
where $a^\dagger$ and $b^\dagger$ are the creation operators for a photon propagating out of the device at the corresponding output ports. Carrying out the multiplication, the amplitude of the coincidence term $a^\dagger b^\dagger$ is $|r_1||r_2| e^{i\text{arg  } r_1 + i \text{arg } r_2} + |t_1| |t_2| e^{i\text{arg } t_1 + i\text{arg } t_2}$. Thus when $|r_1| = 1/\sqrt{2}$, it follows from the above $|r_2| = |t_1| = |t_2| = 1/\sqrt{2}$, reducing the coincidence amplitude to $(e^{i\text{arg }r_1 + i\text{arg } r_2} + e^{i\text{arg }t_1 + i\text{arg } t_2})/2$. When the phase constraint is placed into this expression, it nullifies, obtaining the standard HOM cancellation of coincident terms.

\section{Y-coupler unitary transformation}

\begin{figure}[ht]
    \centering
    \includegraphics[width=.5\textwidth]{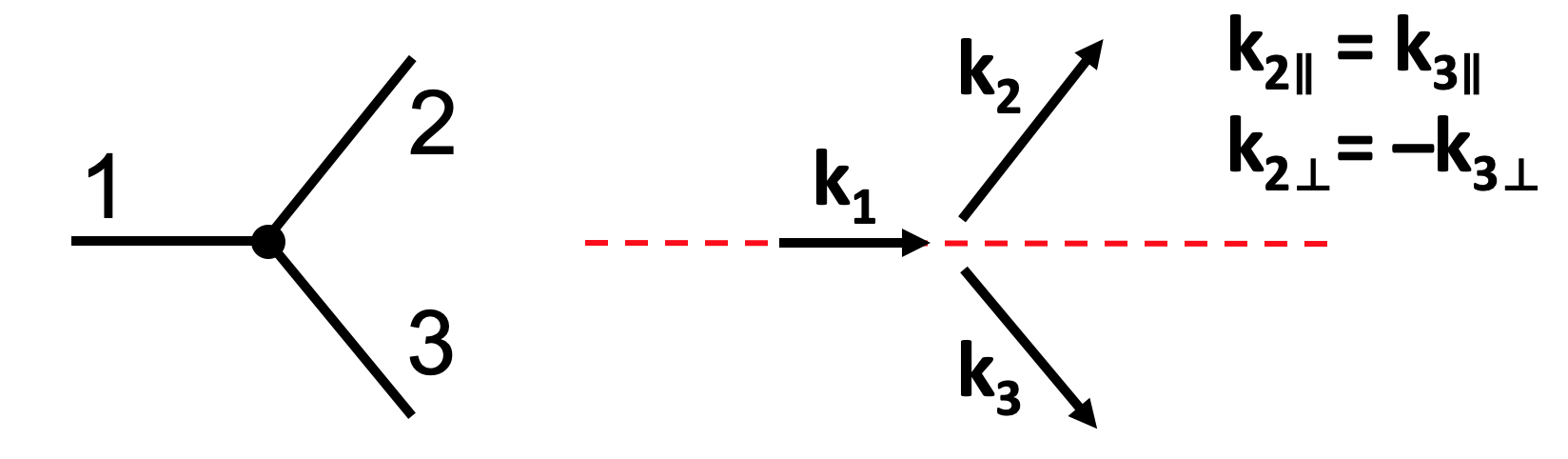}
    \caption{(left) Pictorial depiction of a generic 3-port scatterer, such as the Y-coupler. (right) Illustration of a feed-forward and reflection symmetries with respect to input at port 1, which are the defining properties of the standard Y-coupler. Feed-forward implies a photon is unable to back-reflect and populate the spatial mode with propagation vector $-\mathbf{k_1}$. Meanwhile, the reflection symmetry forces the output mode vectors $\mathbf{k_2}$ and $\mathbf{k_3}$ to be mirrored about the red symmetry line, defined by the direction of the input field. Accordingly, the scattering amplitudes at output ports 2 and 3 are equal.}
    \label{fig:yc}
\end{figure}

A symmetric Y-coupler transformation can be uniquely defined (up to external phase shifts) by three assumptions. Here the term symmetric is used to imply the feed-forward behavior is a 50:50 power splitting from port 1 into ports 2 and 3, and results from a reflection symmetry about the $\mathbf{k}$-vector that defines the propagation direction of the incident plane wave into port 1, which is depicted in Fig. \ref{fig:yc}.  Generalizing the analysis that follows to non-equal splitting ratios is straightforward and is conducted in the appendix, along with an analysis for the case of broken feed-forward symmetry. The assumptions used to define the device are as follows:
\begin{enumerate}
\item The scattering transformation can be represented by a $3 \times 3$ unitary matrix $U$ where $U^\dagger U = 1$. 

\item The scattering process is reciprocal, i.e. $U = U^T$.

\item The device undergoes a feed-forward and reflection-symmetric transformation with respect to input directed at port 1. This can be represented by the transformation
\begin{equation}
|\psi_0\rangle = a_1^\dagger|0\rangle \rightarrow \frac{1}{\sqrt{2}}(a_2^\dagger + a_3^\dagger)|0\rangle.
\end{equation}
\end{enumerate}

The final assumption offers more than the equal intensity, equal phase transformation above; the collection of assumed symmetries also imply that light input to port 2 will undergo the same transformation as port 3 except with the output amplitudes for ports 2 and 3 swapped.

Combining assumptions 2 and 3 allows us to begin with the matrix transformation
\begin{equation}
Y = 
\begin{pmatrix}
0 & 1/\sqrt{2} & 1/\sqrt{2}\\
1/\sqrt{2} & a & b\\
1/\sqrt{2} & b & a
\end{pmatrix}.
\end{equation}
Next, assumption 1 can be enforced by requiring the columns of $Y$ form an orthonormal basis of $\mathbb{C}^3$. In particular, 
\begin{subequations}
\begin{align}
    |a|^2 + |b|^2 &= 1/2\\
    (a + b) &= 0\\
    1/2 + (a^* b + b^*a) &= 0
\end{align}
\end{subequations}
The middle equation implies $a$ and $b$ are offset in phase by $\pi$ and that $|a| = |b|$. Inserting this into the top equation results in the result $|a| = |b| = 1/2$. Next, if we let $a = e^{i\phi_a}/2$ and $b = e^{i\phi_b}/2$, the third equation can be written $\cos (\phi_a - \phi_b) = -1$, confirming the phase relationship between $a$ and $b$ present in the second equation.

Thus, the derived unitary matrix of the symmetric Y-coupler can be written
\begin{equation}\label{eq:yc}
  U =
  \begin{pmatrix}
    0 & 1/\sqrt{2} & 1/\sqrt{2}\\
    1/\sqrt{2} & -1/2 & 1/2\\
    1/\sqrt{2} & 1/2 & -1/2
  \end{pmatrix}
\end{equation}
Here we have chosen the standard phase convention where the reflected field acquires a phase of $\pi$. Other valid phase conventions are easily obtained if external phase shifts are placed outside the scatterer. In this case, the aggregate scattering matrix would become
\begin{equation}
\begin{pmatrix}
0 & e^{i\phi_1 + i\phi_2}/\sqrt{2} & e^{i\phi_1 + i\phi_3}/\sqrt{2}\\
e^{i\phi_1 + i\phi_2}/\sqrt{2} & -e^{2i\phi_2}/2 & e^{i\phi_2 + i\phi_3}/2\\
e^{i\phi_1 + i\phi_3}/\sqrt{2} & e^{i\phi_2 + i\phi_3}/2 & -e^{2i\phi_3}/2 
\end{pmatrix}
\end{equation}
Let $\phi_2 = \phi_3 \coloneqq \phi/2$ to preserve the reflection symmetry. Then fixing $\phi_1 = 2\pi - \phi/2$ gives the matrix
\begin{equation}\label{eq:yc-gen}
\begin{pmatrix}
0 & 1/\sqrt{2} & 1/\sqrt{2}\\
1/\sqrt{2} & -e^{i\phi}/2 & e^{i\phi}/2\\
1/\sqrt{2} & e^{i\phi}/2 & -e^{i\phi}/2
\end{pmatrix}.
\end{equation}
From this we see the choice of phase convention is arbitrary since it can always be changed with external phase shifts. A similar result occurs with a 50:50 beam-splitter, allowing the Hadamard and equal-phase versions to be interchanged with external phase shifts. Thus, replacing one convention with another can in general only translate the system output in its phase parameter space. The parametric curves themselves are the same. 

Equation (\ref{eq:yc}) for the Y-coupler represents the action of a rather irregular optical multiport: when input to port 1 it is directionally-biased (no reflection back), and from ports 2 and 3 it is directionally-unbiased. This hybrid nature allows interesting devices to be constructed, mainly because the orientation of the feed-forward and directionally-unbiased ports can be used to control the appearance of optical resonators. We study some of the simplest devices in the next section.

\begin{figure}[ht]
    \centering
    \includegraphics[width=.45\textwidth]{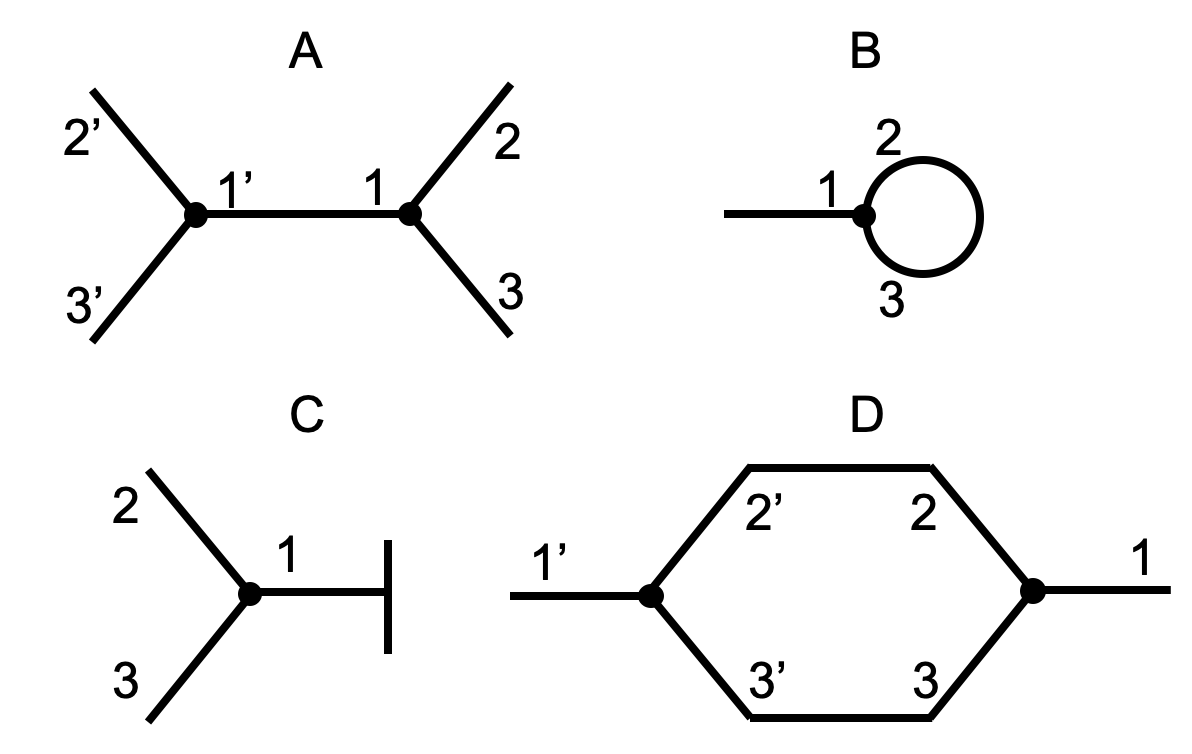}
    \caption{Simple scattering configurations formed from one or two symmetric Y-couplers. The back-reflecting ports labeled 2 and 3 allow interesting devices to be formed: (A) resource-efficient, cavity-free Grover four-port, (B) phase-independent loop mirror, (C) Single-parameter Michelson interferometer, and (D) a tunable resonator that blends together the traditional Fabry-Perot and Mach-Zehnder interferometers.}
    \label{fig:multi}
\end{figure}
\section{directionally-unbiased Y-coupler devices}
\subsection{Four-port Grover coin}

The previously known implementation of the four-port Grover coin is a low-finesse ring-cavity system that requires four mirrors, four beam-splitters, and configuring eight phase parameters \cite{PhysRevA.93.043845}. However, with two Y-couplers, the same device can be realized without causing a resonator to form. In fact, the device does not exhibit any interference whatsoever, so in the absence of other devices, it will be unaffected by phase noise. These properties make the following implementation significantly more stable and resource-efficient than its predecessor.

To obtain the Grover four-port, connect the feed-forward ports of the two Y-couplers together as shown in Fig. (\ref{fig:multi}A). Assume the bridge carries a total phase shift of $\phi$. Then, inputting to any port, there is a back-reflected amplitude of $-1/2$, a transmission amplitude of $1/2$ to the neighboring port, and a transmission into the bridge with amplitude $1/\sqrt{2}$. This portion of the optical state acquires the phase factor $e^{i\phi}$ and then splits equally at the other two ports, each ending with a final scattering amplitude $e^{i\phi}/2$. Since the input port was arbitrary, this result holds for all inputs. The matrix form of this configuration is then
\begin{equation}\label{eq:grovergen}
  U = \frac{1}{2}
  \begin{pmatrix}
    -1 & 1 & e^{i\phi} & e^{i\phi}\\
    1 & -1 & e^{i\phi} & e^{i\phi}\\
    e^{i\phi} & e^{i\phi} & -1 & 1\\
    e^{i\phi} & e^{i\phi} & 1 & -1
  \end{pmatrix}. 
\end{equation}
Any device enacting this transformation will be called a ``generalized Grover four-port'' since the standard four-port Grover coin is matched exactly when $\phi$ is made 0 (modulo $2\pi$). In some applications, allowing $\phi$ to be nonzero provides a valuable degree of freedom that cannot be accessed with external phase shifts.


\subsection{Loop mirror}
A symmetric Y-coupler can also be used to construct a peculiar loop mirror (Fig. \ref{fig:multi}B). This principle has been used extensively in integrated-photonic Sagnac loop mirrors; for a review see \cite{10.1063/5.0123236}. By connecting the directionally-unbiased ports together, the device reflects light entering the loop regardless of the phase it acquires inside the loop. Usually, the phase shift is equally acquired by the two counter-propagating loop modes, under the common assumption of reciprocity. In this case, the loop phase can be factored out as a global phase shift. In general, under ideal operation, this device can only affect the phase of light exiting the device. A resonator will not form unless the phase shift is non-reciprocal. If so, the phase response will be nonlinear in the non-reciprocal phase portion, opening up the possibility of using this device as an enhanced Sagnac phase readout.

\subsection{Michelson interferometer}
When a mirror closes the feed-forward port, the device behaves identically to a Michelson interferometer except now it only depends on a single controllable phase instead of two (Fig. \ref{fig:multi}C). In a standard Michelson, the second phase shift is redundant, since the output probabilities only depend on the difference of the arm phases.

\subsection{2-port tunable resonator}
The final device presented in this paper is a tunable resonator, formed by connecting each pair of directionally-unbiased ports of two symmetric Y-couplers together (Fig. \ref{fig:multi}D). This 2-port device can be viewed as a mixture of a traditional Fabry-Perot and Mach-Zehnder interferometer. Let the top arm that connects modes 2 and 2' in Fig. \ref{fig:multi}D carry a phase shift $\phi_1$. Similarly assume the bottom arm linking modes 3 and 3' carries a phase shift of $\phi_2$. The output scattering amplitudes can be found by analyzing the coupled-cavity supermodes. The computation, detailed in the appendix, is very much like that of the Grover-Michelson interferometer \cite{PhysRevA.107.052615}, and the output amplitudes have a similar structure. Let
\begin{subequations}\label{eq:BC}
\begin{align}
    B &= \frac12 (e^{i\phi_1} + e^{i\phi_2}) \text{ and }\\
    C &= \frac12 (e^{i\phi_1} - e^{i\phi_2}). 
\end{align}
\end{subequations}
\begin{figure}[ht]
    \centering
    \includegraphics[width=.5\textwidth]{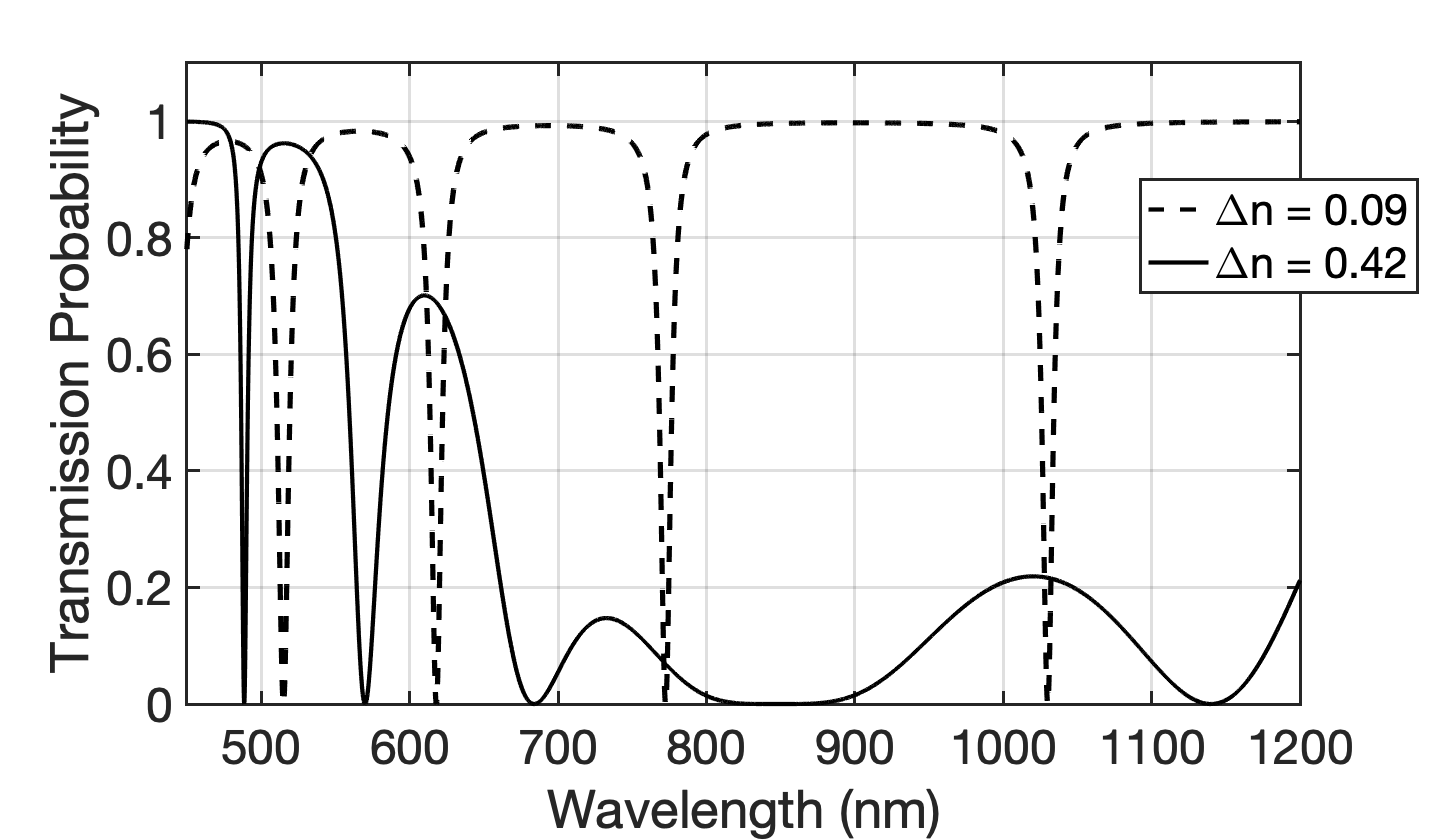}
    \caption{Basic illustration of switchable dichroic behavior. If the two cavity arms are equal length, the device will transmit all wavelengths. With a small index perturbation in one arm (dashed curve), transmission notches appear. A short arm length causes these spectral notches to be widely spaced apart. For a larger index perturbation, the spectral response around any desired region can be made to invert (solid curve) allowing the system to behave as a switchable dichroic mirror. This switching capability is enabled by the additional degree of freedom that the Y-coupler etalon possesses relative to a standard Fabry-Perot.}
    \label{fig:ymzifpi}
\end{figure}
Then for the present device, it can be shown
\begin{subequations}
\begin{align}
    r &= -\frac{C^2}{1 - B^2}\\
    t &= B(1-r).
\end{align}
\end{subequations}

The reflection amplitude $r$ describes how the optical field scatters between the same input and output, namely from port 1 to port 1 or from port 1' to port 1' for the labeling in Fig. \ref{fig:multi}D. $t$ describes the scattering behavior for the light passing through the device, from port 1 to port 1' or vice versa. From these scattering amplitudes $r$ and $t$, the reflection and transmission probability can be found by taking the square modulus: $T(\phi_1, \phi_2) = |t(\phi_1, \phi_2)|^2$ and $R(\phi_1, \phi_2) = |r(\phi_1, \phi_2)|^2$. 

In a traditional Fabry-Perot interferometer or etalon, the finesse is a fixed function of the reflectivity of each optical surface. With high finesse, the system will exhibit narrow transmittance peaks. Here the effective finesse can be controlled by varying the relative phase between the two arms, and the system will now exhibit narrow transmittance dips instead of peaks. When the phase difference between the two arms is varied further, the dips widen, allowing transmission at a particular band of frequencies to be actively switched. 

A basic illustration of this behavior is shown in Fig. \ref{fig:ymzifpi}. When each resonator arm is equal in length, the system fully transmits at each wavelength. Spectral dips form when one phase is perturbed slightly, for instance with a path-length change inducing $\Delta \phi(k) = k n (\Delta \ell)$ or a refractive index change inducing $\Delta \phi(k) = k (\Delta n) \ell$. For a small resonator length, these spectral dips are widely spaced, and the width of each notch be widened by increasing the phase difference between the two arms. Thus at a larger phase change, the system reflects relatively flat bands of frequencies. This allows the system to act as a switchable dichroic. 

\section{Discussion}
\begin{figure}[ht]
    \centering
    \includegraphics[width=.5\textwidth]{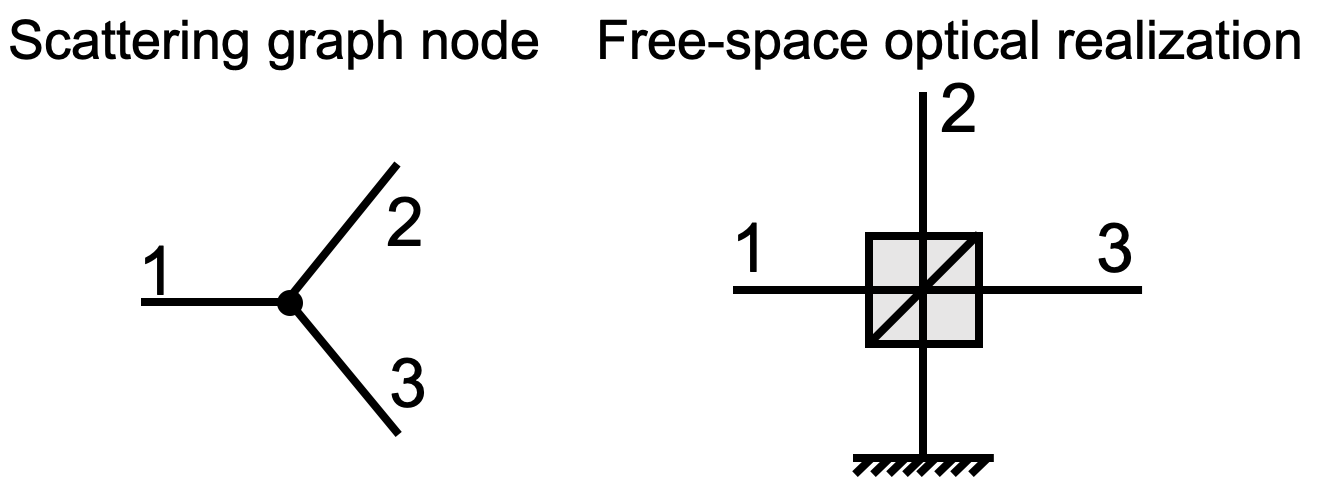}
    \caption{A Y-coupler scattering graph node (left) and a simple free-space optical implementation (right). The splitting-ratio and spectral response is determined by that of the beam-splitter and mirror. A non-equal splitting ratio results in the asymmetric behavior detailed in the appendix. The phase acquired inside the mirror-arm is arbitrary, since it can be equivalently adjusted by external phase shifts. Such external phase shifts merely represent an immeasurable global translation to the value of any controllable phases in the system.}
    \label{fig:yc2}
\end{figure}

A vast number of Y-coupler designs exist in the literature, having the immediate potential to realize many interesting applications of unbiased linear-optical scatterers. In addition to the devices discussed above, a number of similar unbiased devices could be considered by orienting the feed-forward port of the Y-coupler in different ways or by combining the Y-coupler with other multiport scatterers. In addition to the many photonic and fiber implementations of the Y-coupler, it is possible to obtain a Y-coupler transformation in bulk optics with a single beam-splitter. As is suggested from the Y-coupler-based Michelson interferometer discussed above, forming a standard Michelson and removing one end-mirror can be shown to exactly reproduce the transformation of eq. (\ref{eq:yc-gen}). The spectral properties of the beam-splitter and remaining end-mirror determine those of the formed Y-coupler. This free-space configuration is depicted in Fig. \ref{fig:yc2}. Note the phase acquired inside the mirror-arm is arbitrary, since it can be equivalently adjusted by external phase shifts. Such external phase shifts merely represent an immeasurable global translation to the value of any phases around the device when it is placed in a larger system. In accord with this is the following fact: naturally if an additional end-mirror is added to this Y-coupler the system becomes a Michelson interferometer. Changing the first end-mirror's position only translates the sinusoidal interference pattern that is produced by dithering the new end-mirror. 

\begin{figure}[ht]
    \centering
    \includegraphics[width=.5\textwidth]{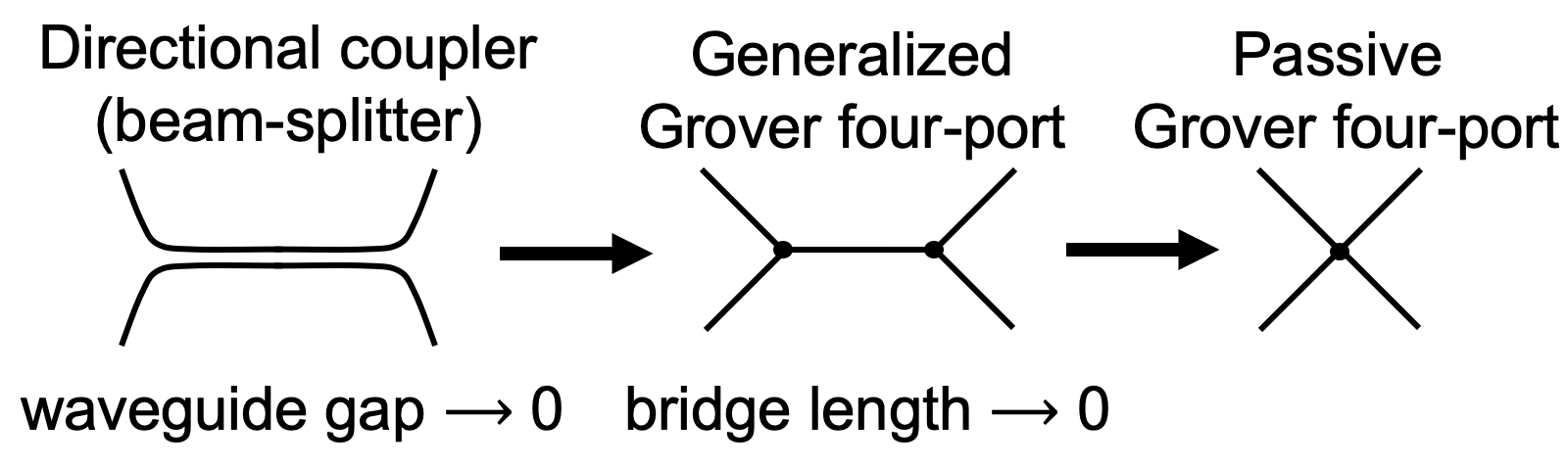}
    \caption{The Y-coupler Grover four-port of eq. (\ref{eq:grovergen}) can be viewed as the device that emerges as the gap of a directional coupler is brought to zero. In practice, the waveguide and coupler parameters must be chosen so that the port created by the fused waveguides does not have back-reflections when light is input to it. If this is so, bringing the bridge length to zero will result in a passive Grover four-port.}
    \label{fig:limit}
\end{figure}

An interesting perspective of the Grover four-port constructed above can be viewed as the result of two limiting processes being applied to a beam-splitter. First, a directional coupler beam-splitter is shown in Fig. (\ref{fig:limit}) (left). The generalized Grover four-port of eq. (\ref{eq:grovergen}) is obtained when the distance between the parallel waveguides is brought to zero, as in Fig. (\ref{fig:limit}) (center). Then, the true Grover coin is obtained in the limit that the middle branch length is brought to zero (right). In practice, if the source coherence length is greater than the branch length, the device will act as a Grover coin as long as the branch phase is an integer multiple of $2\pi$. This is an advantage over the traditional, resonator-based version \cite{PhysRevA.93.043845, PhysRevA.101.032118}, which requires the coherence length to be larger than several round-trip lengths, allowing a greatly improved design to be employed for the Grover coin's growing number of applications, such as higher-dimensional interferometry \cite{PhysRevA.106.033706, PhysRevA.107.052615}, quantum state routing with a multi-mode version of the Hong-Ou-Mandel effect \cite{PhysRevA.104.012617, PhysRevA.104.012617}, and applications of quantum walks such as Hamiltonian simulation, quantum state transfer, and quantum search \cite{PhysRevA.101.032118, PhysRevA.95.042109, PhysRevA.96.013858, kubota2022perfect, giri2017review}.

It remains to be seen if the present approach can be generalized to obtain Grover coins in other dimensions without internal interference. Recursing the structure of Fig. (\ref{fig:multi}A) can readily generate an eight-port Grover coin but requires internal two-beam interference. The present design of the four-port Grover coin might also enable more practical application of the method mentioned in Ref. \cite{PhysRevA.107.052615}: a $(d+m-2)$-port Grover coin can constructed by connecting a $d$-port and $m$-port Grover coin with a bridge. The bridge must carry a phase zero modulo $2\pi$. Similarly, a $(d-m)$-port Grover coin can be obtained by placing $m$ mirror seals on a $d$-port Grover coin, each with round-trip phase of zero modulo $2\pi$. At other values of these phases, the device spans the set of unitary, reciprocal, and circulant $(d-m)$-port scatterers, as discussed in Section II.

The tunable resonator shown in Fig. \ref{fig:multi}D has more degrees of freedom than a traditional Fabry-Perot. With proper control of its phase parameters, a switchable dichroic mirror or bandpass filter could be constructed. When the resonator length increases, a dense transmission-dip comb spectrum appears, which may offer improvements to traditional electro-optic frequency comb systems \cite{Parriaux:20} formed from Fabry-Perot systems. Of course, with multiple frequencies, additional considerations must be taken into account, such as the spectral response of the 3-ports used to form the device and dispersion.

In this work, we derived a unitary scattering transformation for a Y-coupler from symmetry considerations. The device is somewhat unconventional, being a passive scatterer which is a hybrid of feed-forward and back-reflecting. We have presented but a small fraction of the possible Y-coupler configurations. The hybrid scattering nature allows control over whether resonators do or do not form in larger systems, and thus might make an interesting centerpiece for study of quantum walks on a 3-regular lattice. It also remains to be seen how symmetry might be used to find other useful optical scatterers, and what benefits the endowed symmetry provides.

\section*{Acknowledgements}
This research was supported by the Air Force Office of Scientific Research MURI Award No. FA9550-22-1-0312.

\section*{Appendix I: Y-coupler generalizations}
In single-photon scattering processes it is typical to assume unitarity and reciprocity. Arriving at a unique, single-photon description of the symmetric Y-coupler requires two auxiliary assumptions regarding how the device acts on light entering port 1. First, input to port 1 is feed-forward, and second, the outputs at port 2 and 3 were assumed to have a reflection symmetry for this input. Here we provide a separate description of a Y-coupler device that is valid when each of these symmetries break. The reciprocity assumption could also be lifted to further generalize the above cases using a similar approach. This results in devices that can be decomposed with a circulator but will be considered in detail elsewhere. Abandoning both assumptions above would cause the device to lack any characteristic of a Y-coupler, and thus will also not be considered here.

\subsection{Asymmetric Y-coupler}

Relaxing the condition of a reflection-symmetric port 1 gives the matrix
\begin{equation}\label{eq:asym}
Y = 
\begin{pmatrix}
0 & t_1 & t_2\\
t_1 & a_1 & b\\
t_2 & b & a_2
\end{pmatrix}.
\end{equation}
We again enforce unitarity and conduct a similar analysis to the above. Assuming $|t_1| \in (0, 1)$ results in the following restrictions on the probabilities
\begin{subequations}
\begin{align}
|t_1|^2 + |t_2|^2 &=1\\
|a_1|^2 &= |t_2|^4\\
|a_2|^2 &= |t_1|^4\\
|b|^2 &= |t_1|^2|t_2|^2.
\end{align}
\end{subequations}
Phase constraints are also derived from the orthogonality requirement of the columns of the matrix (\ref{eq:asym}). From the first and second columns, as well as second and third columns, we obtain
\begin{equation}\label{eq:phasec}
\phi_{t_2} - \phi_{t_1} + \pi = \phi_b - \phi_{a_1} = \phi_{a_2} - \phi_b.
\end{equation}
The final orthogonality constraint, obtained from columns two and three, is redundant. Explicitly, we have
\begin{equation}
|t_1||t_2|e^{i\phi_{t_2} - i\phi_{t_1}} + |b|(|a_1| e^{i\phi_b - i\phi_{a_1}} + |a_2| e^{i\phi_{a_2} - i\phi_b}) = 0.
\end{equation}
The factors $e^{i\phi_b - i\phi_{a_1}}$ and $e^{i\phi_{a_2} - i\phi_b}$ can both be replaced with $-e^{i\phi_{t_2} - i\phi_{t_1}}$ by Eq. (\ref{eq:phasec}). After respectively interchanging $|a_1|$ and $|a_2|$ for $|t_2|^2$ and $|t_1|^2$, this constraint becomes
\begin{equation}
    |t_1||t_2|e^{i\phi_{t_2} - i\phi_{t_1}} (1 - |t_1|^2 - |t_2|^2) = 0,
\end{equation}
which is always true by virtue of the fact $|t_1|^2 + |t_2|^2 = 1$.
\begin{widetext}
Combining everything, we let $t \in (0, 1)$ and put
\begin{equation}
Y = 
\begin{pmatrix}
0 & t e^{i\phi_{t_1}} & e^{i\phi_{t_2}} \sqrt{1 - t^2}\\
t e^{i\phi_{t_1}}  & -(1-t^2)e^{i\phi_b - i(\phi_{t_2} - \phi_{t_1})} & \sqrt{t^2 (1-t^2)} e^{i\phi_{b}} \\
e^{i\phi_{t_2}} \sqrt{1 - t^2} & \sqrt{t^2 (1-t^2)} e^{i\phi_{b}} & -t^2 e^{\phi_b + i(\phi_{t_2} - \phi_{t_1})}
\end{pmatrix}
\end{equation}
where $\phi_{a_1}$ and $\phi_{a_2}$ have been removed using the above phase constraints. If we then define $\Delta = \phi_{t_2} - \phi_{t_1}$ and add external phases of $\phi_1 = -\phi_{t_1}$, $\phi_2 = \phi_3 = -\phi_b$, a global factor $e^{-i\phi_b}$ emerges. This only works because the top-left entry of the above matrix is zero. After discarding the global phase factor, we are left with the two-parameter device
\begin{equation}
Y = 
\begin{pmatrix}
0 & t & e^{i\Delta} \sqrt{1 - t^2}\\
t & -(1-t^2)e^{- i\Delta} & \sqrt{t^2 (1-t^2)} \\
e^{i\Delta} \sqrt{1 - t^2} & \sqrt{t^2 (1-t^2)} & -t^2 e^{i\Delta}
\end{pmatrix}.
\end{equation}
\end{widetext}
Therefore, relaxation of the reflection symmetry is encapsulated by $t$ and $\Delta$. If $t \rightarrow 1/\sqrt{2}$ and $\Delta \rightarrow 0$ the symmetric Y-coupler emerges. 

The asymmetric scattering transformation found here can also be realized in bulk optics by placing a mirror at one port of a unitary beam-splitter with asymmetric scattering coefficients.

\subsection{Directionally-unbiased symmetric Y-coupler}
Removing the requirement of a feed-forward port 1 gives the matrix
\begin{equation}
Y = 
\begin{pmatrix}\label{eq:ycgen1}
r & t & t\\
t & a & b\\
t & b & a
\end{pmatrix}.
\end{equation}
Enforcing unitary produces four equations
\begin{subequations}
\begin{align}
    |r|^2 + 2|t|^2 &= 1,\label{eq:r}\\
    |t|^2 + |a|^2 + |b|^2 &= 1,\label{eq:t}\\
    |t|^2 + (a^*b + b^*a) &= 0,\label{eq:orth} \text{ and }\\
    r^* t + t^*(a + b) &= 0\label{eq:orth2}.
\end{align}
\end{subequations}

The matrix (\ref{eq:ycgen1}) has four different complex-valued entries, i.e. four magnitudes and four phases. The four equations above can remove four of the eight degrees of freedom. With three external phases we can remove only two additional parameters, because the phases at ports 2 and 3 must be the same to maintain the assumed reflection symmetry.

Thus we expect that up to external phase shifts the present device can be described by two free parameters. This is consistent with the fact that only the variable $r$ has been introduced with this generalization. When $r$ is brought to zero, the phase $r$ carries loses its relevance, so two parameters are removed.

Rewriting Eq. (\ref{eq:orth}), we have $|t|^2 + 2 |a||b|\cos(\phi_a - \phi_b) = 0$. Since $|t|^2$ and $2|a||b|$ are both nonnegative, this equation is solvable only for $x = \cos(\phi_a - \phi_b) \leq 0$, with the case $x = 0$ corresponding to $|t| = 0$. Replacing $|t|^2$ with $1 - |a|^2 - |b|^2$ we to obtain
\begin{equation}
     |a|^2 - 2|a||b| x + |b|^2 = 1.
\end{equation}
This is an elliptical equation for $|a|$ and $|b|$, with $x$ affecting the eccentricity. We can use this equation to find $|a|$ in terms of $|b|$ or vice versa. Solving for $|b|$ we have 
\begin{equation}\label{eq:b}
    |b| = |a| x \pm \sqrt{|a|^2 (x^2 - 1) + 1}.
\end{equation}
Because $x \leq 0$, we must take the positive root, otherwise $|b|$ becomes negative. From this equation, a provided value of $\leq 0 |a| \leq 1$ and $-1 \leq x \leq 0$ can be used to find $|b|$. Then $|a|$ and $|b|$ dictate $|t|$ through Eq. (\ref{eq:t}), which in turn dictates $|r|$ through Eq. (\ref{eq:r}).

The remaining unknown is the phase between $r$ and $t$, which we can obtain from the orthogonality constraint Eq. (\ref{eq:orth2}). To simplify what follows without any loss of generality, we now use the external phase freedom to fix $\phi_b = 0$ and $\phi_r = 0$. The necessary external phases required to obtain this are $\phi_2 = \phi_3 = -\phi_b/2$ and $\phi_1 = -\phi_r/2$; this transformation will translate $\phi_t$ and $\phi_a$ but this can be absorbed into their definition.

Omitting the trivial case $|t| \neq 0$, Eq. (\ref{eq:orth2}) can be rewritten as
\begin{equation}
    |r|e^{2i\phi_t} + |a|e^{i\phi_a} + |b| = 0.
\end{equation}
Now we define $\Delta \coloneqq 2\phi_t$ and remove $|b|$ using Eq. (\ref{eq:b}), which gives
\begin{equation}
    |r|e^{i\Delta} + |a|e^{i\phi_a} + |a| x + \sqrt{|a|^2 (x^2 - 1) + 1} = 0,
\end{equation}
or split into real and imaginary parts, 
\begin{subequations}
\begin{align}
    |r|\sin\Delta + |a|\sin\phi_a = 0, \text{ and } \label{eq:top} &\\
    |r|\cos\Delta + 2|a| x + \sqrt{|a|^2 (x^2 - 1) + 1} = 0\label{eq:bottom}.
\end{align}
\end{subequations}
Since $|r| = 0$ was already considered in the main text, we assume otherwise and put $\sin\Delta = -|a|\sin\phi_a/|r| = -(|a|/|r|)\sqrt{1-x^2}$.

Now generally $\sin(\pi - \Delta) = \sin(\Delta)$ but $\cos(\pi - \Delta) = -\cos(\Delta)$. Thus we always have the freedom to invert the sign of $\cos\Delta$ without affecting $\sin\Delta$. This enables us to write $\cos\Delta = \pm \sqrt{1 - \sin^2\Delta} = \pm \sqrt{1 - |a|^2/|r|^2 (1-x^2)}$ with the understanding that the leading sign may be picked freely without disturbing $\sin\Delta$, preserving the constraint of Eq. (\ref{eq:top}). Placing this equation into the bottom equation (\ref{eq:bottom}), we have 
\begin{equation}\label{eq:lm}
\pm \sqrt{|r|^2 - |a|^2 (1-x^2)} = 2|a|x + \sqrt{|a|^2(x^2 - 1) + 1}.
\end{equation}


The final step is to show that Eq. (\ref{eq:lm}) is valid for any choice of $-1 \leq x \leq 0 $ and $0 \leq |a| \leq 1$. Choosing $\Delta$ so that each side has the same sign, we only need to compare their magnitudes, or equivalently, their magnitudes squared:
\begin{multline}\label{eq:long}
    |r|^2 - |a|^2(1-x)^2 = 4|a|^2x^2 + |a|^2x^2 \\ - |a|^2 + 1 + 4|a|x\sqrt{|a|^2(x^2-1)+1}. 
\end{multline}
Or, after cancelling terms, 
\begin{equation}\label{eq:rsq}
    |r|^2 = 4|a|^2x^2 + 1 + 4|a|x\sqrt{|a|^2(x^2-1)+ 1}
\end{equation}
Now we replace $|r|^2$ with an independently derived expression. We have that 
\begin{equation}
    |a|^2 + |b|^2 = |a|^2 + \textbf{(}|a|x + \sqrt{|a|^2(x^2-1) + 1}\textbf{)}^2
\end{equation}
while
\begin{subequations}
\begin{align}
    |r|^2 &= 1 - 2|t|^2 = 2(|a|^2 + |b|^2) - 1 \\ &= 2|a|^2 + 2\textbf{(}|a|^2x^2 + |a|^2(x^2 - 1) + 1  \\ 
    \notag & + 2|a|x\sqrt{|a|^2(x^2-1) + 1} \textbf{)} -1 \\ &= 
    4|a|^2x^2 + 1 + 4|a|x\sqrt{|a|^2(x^2-1)} \label{eq:last}.
\end{align}
\end{subequations}
Hence, we see Eq. (\ref{eq:last}) and Eq. (\ref{eq:rsq}) are in fact identical, indicating that for this choice of $\Delta$, Eq. (\ref{eq:orth2}) is satisfied with no additional restrictions imposed between $|a|$ and $x$. Thus $|a|$ and $x$ are free parameters describing this class of back-reflecting symmetric Y-couplers, with all others being computed directly from these.

Although we omitted the case $|r| = 0$ in deriving $\Delta$ above, we see in the limit that when $|r|$ goes to zero $\Delta$ gets removed from the equations (\ref{eq:top}) and (\ref{eq:bottom}). This is consistent with the fact that a zero-energy back-reflection can be assigned any phase, allowing the external phase freedom at port 1 to be used to nullify $\Delta$. 

In the special case $x= -1$, we have an explicit parametrization in terms of $|r|$ alone. $x = -1$ implies $|a| + |b| = 1$ with $\phi_a = \pi$, so that $a = -|a|$ and $b = |b|$. This can be used in conjunction with Eqs. (\ref{eq:r}) and (\ref{eq:t}) to obtain a quadratic in either $|a|$ or $|b|$ that depends on $|r|^2$. If $q$ is either $|a|$ or $|b|$, the quadratic is
\begin{equation}
4q^2 - 4q + (1-|r|^2) = 0,
\end{equation}
which is solved for $q = (1 \pm |r|)/2$. Both solutions are valid, but the signs must be selected oppositely for $a$ and $b$ in order for $|a| + |b| = 1$ to remain true. 

Because there are two ways to choose the signs, the form of the device can be split into two classes. In the first class, $|a| = (1 + |r|)/2$ and $|b| = (1-|r|)/2$ so that when light is input input to port two (three), more of it reflects back than transmits to port three (two). In the other class, the situation is reversed: $|a| = (1 - |r|)/2$ and $|b| = (1+|r|)/2$, so that less light reflects. 

Combining $r^* t + t^* (a + b) = 0$ with the above gives
\begin{equation}
    |t||r|( e^{i\phi_t - i\phi_r} \mp e^{-i\phi_t}) = 0.
\end{equation}
We assume $|t|$ and $|r|$ are nonzero, otherwise, the device behaves according to the above or like a mirror. After dividing these quantities out, we have
\begin{equation}
    e^{2i\phi_t - i\phi_r} \mp 1 = 0
\end{equation}
so that in the first class we have $\phi_t = \phi_r/2$ and in the second class $\phi_t = (\phi_r + \pi)/2$. If we use the external phase freedom at port 1 to set $\phi_r = 0$, we can now write out a specific matrix for both classes of the Y-couplers considered here. We have, for real $r \in (0, 1),$
\begin{equation}
    Y_{\pm} = 
\begin{pmatrix}
    r & \sqrt{(1 - r^2)/2} & \sqrt{(1 - r^2)/2}\\
    \sqrt{(1 - r^2)/2} & -(1 \pm r)/2 & (1 \mp r)/2 \\ 
    \sqrt{(1 - r^2)/2} & (1 \mp r)/2 & -(1 \pm r)/2
\end{pmatrix}.
\end{equation}
Alternatively, one could combine these into a single matrix by allowing $r$ to be negative, viz. to let $r \in (-1, 1)$. In any case, taking $r \rightarrow 0$ reproduces the result in the main text. This generalization of the device can be realized by placing a two-port unitary scatterer with reflection amplitude $r$ in port 1 of the symmetric, feed-forward Y-coupler.

\section*{Appendix II: 2-port Y-coupler interferometer output}
We illustrate the output computation for the device shown in Fig. \ref{fig:multi}D for one of the two input ports. The other input behavior is the same due to the underlying symmetry. 

First, we identify photon creation operators with the different spatial modes that are coupled by the device. With reference to Fig. \ref{fig:multi}D, port 1' will be identified with $a^\dagger$, Port 1 with $b^\dagger$, the 2-2' arm with $a_1^\dagger$ and the 3-3' arm with $a_2^\dagger$. We recall the 2-2' arm carries phase $\phi_1$ and the 3-3' arm carries $\phi_2$. 

Next, we show that the quantity $(a_1^\dagger - a_2^\dagger)$ is a supermode of the internal cavity. $L$ and $R$ subscripts will be used denote leftward and rightward propagation directions inside this cavity. We have, during a single round-trip,
\begin{align}
    (a_1^\dagger - a_2^\dagger)_L &\rightarrow \frac{1}{\sqrt{2}} (e^{i\phi_1} - e^{i\phi_2}) a^\dagger \\ \notag &- \frac12 (e^{i\phi_1} + e^{i\phi_2}) (a_1^\dagger - a_2^\dagger)_R \\ &= \sqrt{2} C a^\dagger - B(a_1^\dagger - a_2^\dagger)_R \\&\rightarrow \sqrt{2} C a^\dagger - \sqrt{2} BC b^\dagger + B^2 (a_1^\dagger - a_2^\dagger)_L.
\end{align} 
$B$ and $C$ are functions of $\phi_1$ and $\phi_2$ defined in Eq. \ref{eq:BC}. The direct recursion of each round-trip shown in the preceding equation is a manifestation of the fact that this superposition of the cavity modes $a_1^\dagger$ and $a_2^\dagger$ is an eigenmode (or supermode) of the coupled-mode resonator. The coefficients preceding the operators $a^\dagger$ and $b^\dagger$ are readily turned into a geometric series in the round-trip number $T$ and summed to obtain the steady-state output:
\begin{equation}\label{eq:ss}
(a_1^\dagger - a_2^\dagger)_L \xrightarrow[]{T\rightarrow \infty} \sqrt{2} \bigg (\frac{C}{1-B^2} a^\dagger - \frac{BC}{1-B^2} b^\dagger\bigg)
\end{equation}
Now we start with our initial state $|\psi_0\rangle = a^\dagger|0\rangle$, which transiently (before any round-trips) maps to
\begin{equation}
    a^\dagger|0\rangle \rightarrow \bigg (B b^\dagger - \frac{1}{\sqrt{2}} C(a_1^\dagger - a_2^\dagger)_L \bigg )|0\rangle.
\end{equation}
Taking $T\rightarrow \infty$ and substituting in the result of Eq. (\ref{eq:ss}), we find the output state 
\begin{equation}
    |\psi_{\text{out}}\rangle = \bigg ( B b^\dagger - C \bigg [ \frac{C}{1-B^2}a^\dagger - \frac{BC}{1-B^2} b^\dagger \bigg ] \bigg ) |0\rangle
\end{equation}
from which we extract 
\begin{equation}
r = -\frac{C^2}{1-B^2} \text{, } t = B\bigg (1 + \frac{C^2}{1-B^2}\bigg ) = B(1-r).    
\end{equation}

\bibliography{refs}
\end{document}